\documentclass[a4paper,12pt]{article}

\usepackage[T2A]{fontenc}
\usepackage[utf8]{inputenc}
\usepackage{amsmath}

\numberwithin{equation}{section}

\usepackage[english]{babel}
\usepackage{amsfonts}
\usepackage{epsfig}
\usepackage{booktabs}
\usepackage{bbold}

\newcommand{\bea}{\begin{array}}
\newcommand{\eea}{\end{array}}

\newcommand{\nnr}{\nonumber\\}
\newcommand{\ex}{{\rm e}}

\newcommand{\CC}{{\mathbb C}}
\newcommand{\RR}{{\mathbb R}}

\newcommand{\beq}{\begin{equation}}
\newcommand{\eeq}{\end{equation}}
\newcommand{\fr}{\frac}
\newcommand{\beqn}{\begin{eqnarray}}
\newcommand{\eeqn}{\end{eqnarray}}
\newcommand{\pt}{\partial}

\newcommand{\eps}{\epsilon}

\newcommand{\diag}{{\rm diag}}

\title{}
\author{}
\date{}
\begin{document}
\begin{titlepage}

\hfill\parbox{40mm}
{\begin{flushleft}  ITEP-TH-XX/12
\end{flushleft}}

\vspace{30mm}

\begin{center}
{\large \bf On the Solutions of Generalized Bogomolny Equations}

\vspace{17mm}

\textrm{Victor~Mikhaylov}
\vspace{8mm}

\textit{Physics Department, Princeton University,\\
Princeton, NJ 08540, USA}\\
\texttt{victor.mikhaylov AT gmail.com}

\vspace{3.5cm}

{\bf Abstract}\end{center}
Generalized Bogomolny equations are encountered in the localization of the topological ${\mathcal N}=4$ SYM theory. The boundary conditions for 't~Hooft and surface operators are formulated by giving a model solution with some special singularity. In this note we consider the generalized Bogomolny equations on a half space and construct model solutions for the boundary 't~Hooft and surface operators. It is shown that for the 't~Hooft operator the equations reduce to the open Toda chain for arbitrary simple gauge group. For the surface operators the solutions of interest are rational solutions of a periodic non-abelian Toda system.

\end{titlepage}
\section{Introduction}
The maximally supersymmetric Yang-Mills theory in four dimensions can be twisted in a number of ways to obtain topological field theories. One of the twists appears to be relevant for the geometric Langlands program \cite{Langlands} and for the Chern-Simons topological theory with a non-trivial integration cycle \cite{ComplexCS, 5knots, GaiottoWitten}. In the latter case the four-dimensional theory is defined on a manifold with boundary, so that the Chern-Simons theory is effectively induced on this boundary. For supersymmetric observables the computations localize on solutions of the supersymmetry conditions, which in this case are the generalized Bogomolny equations introduced in \cite{Langlands}. The natural Chern-Simons observables correspond to the boundary 't~Hooft or surface operators in four dimensional language. These operators are defined by prescribing the singular behavior of the fields near the boundary to be similar to the model solutions of the supersymmetry equations.

For the $SU(2)$ gauge group these model solutions were found in sec.~3 of \cite{5knots}. For higher rank groups solutions for the 't~Hooft operator were obtained for special values of the magnetic weight in \cite{Henningson}. In this note we construct model solutions for the boundary 't~Hooft operator with general magnetic weight for any simple compact gauge group. We also find solutions for the surface operator for the $SU(N)$ groups. This turns out to be possible because, as we show, for the boundary 't~Hooft operator the equations reduce to the one-dimensional open Toda chain\footnote{For $SU(N)$ groups this was noted in \cite{note}}, and for the boundary surface operator one gets a kind of non-Abelian Toda equation. Both systems are integrable and thus can be exactly solved. This relation is not particularly surprising. Different reductions of the generalized Bogomolny equations can give, {\it e.g.}, Nahm's equations, Bogomolny equations or Hitchin equations.

The paper is organized as follows. In section~2 a useful reduction of the equations for the case of time independent solutions is reviewed. In section~3 we consider the boundary 't~Hooft operators. In this case the generalized Bogomolny equations can be reduced to an open abelian Toda chain. The problem of finding explicit solutions boils down to solving a system of algebraic equations. We conjecture a formula for the solution for this system, prove this conjecture for the $SU(N)$ groups and also check it for $SO(5)$ and $G_2$. In section~4 the case of surface operators is considered. For this case the equations can be reduced to a version of non-abelian Toda system. We describe how to construct solutions for $SU(N)$ gauge group using the method of Baker-Akhiezer function. Relevant solutions appear to be rational. In the Appendix some explicit formulae for the 't~Hooft operator solution are collected.

It would be interesting to generalize our discussion of surface operators to the case of arbitrary gauge group. 

\section{Reduction of the Bogomolny equations}
Let $G$ be a compact semi-simple Lie group, the corresponding Lie algebra will be denoted by $\mathfrak{g}$. At special value of the twisting parameter the generalized Bogomolny equations are
\beqn
&&F-\phi\wedge\phi+\ast{\rm d}_A\phi=0\,,\label{bogomol1}\\
&&d_A\ast\phi=0\,,\label{bogomol2}
\eeqn
where $A$ is a four-dimensional gauge field, and $\phi$ is a one-form valued in the adjoint of the gauge group $G$. We denote Euclidean time by $x^0$, and the three spacial coordinates by $x^1+ix^2=z$ and $x^3=y$. We will also denote the absolute value of $z$ by $r$.

First we briefly recall the arguments of \cite{5knots} that lead to a useful reformulation of the generalized Bogomolny equations as a moment map for the complex gauge transformations. As explained in \cite{5knots}, for time-independent solutions one can take $A_0=\phi_3=0$, and, introducing the covariant derivatives
\beqn
&&{\mathcal D}_1=2\pt_{\bar z}+A_1+iA_2\,,\nnr
&&{\mathcal D}_2=\pt_y+A_3-i\phi_0\,,\nnr
&&{\mathcal D}_3=\phi_1-i\phi_2\,,
\eeqn
the equations can be reformulated as
\beqn
&&[{\mathcal D}_i,{\mathcal D}_j]=0\,,\quad i,j=1..3\,,\label{commut}\\
&&\sum_{i=1}^3[{\mathcal D}_i,{\mathcal D}_i^\dagger]=0\,.\label{moment}
\eeqn
While the equations (\ref{bogomol1}), (\ref{bogomol2}) are invariant under $G$-valued gauge transformations, the commutativity conditions (\ref{commut}) have a larger symmetry. They are invariant under gauge transformations with values in the complexified gauge group $G_{\CC}$. Then, with a suitable choice of symplectic structure, the remaining equation (\ref{moment}) can be interpreted as a moment map constraint for this symmetry. In components the equation (\ref{moment}) is
\beq
4F_{z\bar z}-2iD_3\phi_0+[\varphi,\varphi^\dagger]=0\,,
\eeq
where we denoted\footnote{In our conventions the fields are antihermitean in unitary representations.} $\varphi=\phi_1-i\phi_2$.

The solution of commutativity conditions (\ref{commut}) is simple. By a gauge transformation the connection in ${\mathcal D}_1$ and ${\mathcal D}_2$ can be gauged away, so we take $A_1=A_2=A_3=0$ and $\phi_0=0$. The remaining equations say that $\varphi$ is holomorphic and independent of $y$. These formulae of \cite{5knots} will be our starting point.

Let $\varphi_0(z)$ be a holomorphic $\mathfrak{g}$-valued function. It defines a solution of the commutativity conditions. We apply a $G_\CC$-valued gauge transformation $g$ to this solution and substitute the resulting fields into the moment map equation (\ref{moment}). The equation becomes
\beq
4\pt_z\left(\pt_{\bar z}h\,h^{-1}\right)+\pt_y\left(\pt_yh\,h^{-1}\right)+[\varphi_0^\dagger(z),h\varphi_0(z)h^{-1}]=0\,,\label{fullmt}
\eeq
where $h=g^\dagger g$. The unitary part of the gauge transformation cancels out in $h$, as it should.

An important case is when the gauge transformation $g$ can be taken to be in the maximal torus of the gauge group. Let $\mathfrak{h}\subset\mathfrak{g}_\CC$ be a real Cartan subalgebra of the split real form of $\mathfrak{g}_\CC$. Then we take $g=\exp(\Psi)$ for $\Psi\in \mathfrak{h}$. In this case the equation (\ref{fullmt}) reduces to
\beq
\Delta_{3d}\Psi+\fr12[\varphi_0^\dagger(z),\ex^{2\Psi}\varphi_0(z)\ex^{-2\Psi}]=0\,.\label{abelmt}
\eeq

\section{Boundary 't~Hooft operators}
An important class of solutions is represented by supersymmetric boundary line defects \cite{5knots}. We work on $\RR^3\times\RR^+$, so $y$ runs from zero to infinity. We choose the boundary 't~Hooft operator to lie along the line $z=0$, $y=0$.
\subsection{Open Toda equations from the Bogomolny equations}
Let $\Delta=\{\alpha_i\}\,, i=1\dots {\rm rank}(\mathfrak{g})$, be the set of simple roots of the simple Lie algebra $\mathfrak{g}_\CC$, and $A_{ij}$ be the Cartan matrix of $\mathfrak{g}_\CC$. We will use the Chevalley basis in the Lie algebra. Denote the raising and lowering operators corresponding to the simple roots by $E^\pm_i$, and the corresponding coroots by $H_i$. Then the commutation relations in the algebra are
\beqn
&&[H_i,H_j]=0\,,\nnr
&&[H_i,E^\pm_j]=\pm A_{ji}E^\pm_j\,,\nnr
&&[E^+_i,E^-_j]=\delta_{ij}H_j\,.
\eeqn
The 't~Hooft operators are labeled by the elements of the cocharacter lattice $\Gamma_{ch}^\vee\in{\mathfrak h}$, up to Weyl equivalence. The cocharacter lattice is just the lattice of homomorphisms ${\rm Hom}\left(\CC^*,G_\CC\right)$. Let $g(z)=z^{\omega}$ with $\omega=\sum_i k_iH_i\in \Gamma_{ch}^\vee$ be such a homomorphism corresponding to our 't~Hooft operator. The Weyl equivalence can be used to transform $\omega$ to the positive Weyl chamber so that
\beq
\mathfrak{r}_i=\alpha_i(\omega)\ge 0\,.
\eeq
The lattice $\Gamma_{ch}^\vee$ lies inside the dual root lattice $\Gamma_r^*$, and hence the numbers $\mathfrak{r}_i$ are integer.

Let $\varphi_1=\sum_i E^+_i$ be a representative of the principal nilpotent orbit in the algebra. We take the solution of the holomorphic equations to be of the form
\beq
\varphi_0(z)=g(z)\varphi_1 g^{-1}(z)\,.
\eeq
Using the commutation relations of the algebra, we have
\beq
\varphi_0(z)=\sum_i z^{\mathfrak{r}_i} E^+_i\,.
\eeq
After a real gauge transformation of this solution $g=\exp(\Psi)$, $\Psi\in\mathfrak{h}$, the fields become
\beqn
&&A_a=-i\eps_{ab}\pt_b\Psi\,,\quad a,b=1..2\,,\nnr
&&\phi_0=-i\pt_y\Psi\,,\quad A_3=0\,,\nnr
&&\varphi=e^\Psi \varphi_0 e^{-\Psi}\,.
\eeqn
A convenient parameterization is $\Psi=\fr12 \sum_{i,j}A^{-1}_{ij}H_i\psi_j$. In this parameterization after some algebra the equation (\ref{abelmt}) gives the following system of equations,
\begin{flalign}
\sum_j A^{-1}_{sj}\Delta_{3d}\psi_j-r^{2\mathfrak{r}_s}\ex^{\psi_s}=0\,.
\end{flalign}
A change of variables
\beq
\psi_i=q_i-2m_i\log r\,,\quad m_i=\mathfrak{r}_i+1\,,
\eeq
brings the equations to the scale invariant form. Assuming scale invariance of the solution, $q_i$ depend only on the ratio $y/r$. It is convenient to introduce a new variable $\sigma$ such that $y/r=\sinh\sigma$. Then the equations take the Toda form\footnote{Note that the sign of the potential terms in these Toda equations is different from the usual one. It is known that in this case the Hamiltonian flows can be incomplete. Indeed, our boundary condition (\ref{bc0}) requires precisely that $\sigma=0$ must be a singular point of the flow.},
\beq
\ddot{q}_i-\sum_j A_{ij}\ex^{q_j}=0\,,\label{Toda}
\eeq
where the dots denote derivatives w.r.t. $\sigma$.

Now, having the equation, we also need the boundary conditions. The boundary condition on the plane $y=0$ away from the defect, {\it i.e.} at $\sigma\to0$, is defined by prescribing the singular behaviour of the fields \cite{Gaiotto,5knots}. In the model solution the gauge field and the normal component of one-form are zero, and the tangent components of the one-form have a singularity
\beq
\phi_0=\fr{t_3}{y}\,, \quad \varphi=\fr{t_1-it_2}{y}\,,\label{principal}
\eeq
where $t_i\in \mathfrak{g}_\CC$ define a principle embedding of the $\mathfrak{su}(2)$ subalgebra. A convenient representative of this conjugacy class is given by
\beqn
&&t_3=\fr{i}{2}\sum_i B_iH_i\,,\nnr
&&t_1-it_2=\sum_i\sqrt{B_i}E^+_i\,,\quad B_i=2\sum_j A^{-1}_{ij}\,.
\eeqn
We note that if $\delta^\vee\in\mathfrak{h}$ is the dual of the Weyl vector in the sense that $\alpha_i(\delta^\vee)=1$, then $t_3=i\delta^\vee$. Also, the numbers $B_i$ which appear in these formulae are in fact positive integers. As they will be important in what follows, we have to explain what they mean (see {\it e.g.} \cite{Cahn}). Let $\Lambda_i$ be a fundamental weight, {\it i.e.} the highest weight in the corresponding fundamental representation. The lowest weight in this representation is of the form $\tilde\Lambda_i=\Lambda_i-\sum_j n_j\alpha_j$. Its level, {\it i.e.} the integer $\sum_j n_j\,$, is called the height of the representation, and it is exactly the number $B_i$.

To rewrite the boundary condition in terms of our ansatz, we note, following \cite{5knots}, that the Toda equations (\ref{Toda}) have a simple solution,
\beq
q_i=-2\log\sinh\sigma+\log B_i\,.\label{model0}
\eeq
The fields on this solution are the following,
\beq
A_a=i\eps_{ab}\fr{x_b}{r^2}\omega\,,\quad \phi_0=\fr{i}{2y}\sum_i B_iH_i\,,\quad \varphi=\fr{1}{y}\sum_i\left(z/\bar z\right)^{\mathfrak{r}_i/2}B_i^{1/2}E^+_i\,.
\eeq
For $\mathfrak{r_i}\ne0$ the solution is singular along the line $r=0$, but a singular unitary gauge transformation with matrix
\beq
\tilde g=\left(\bar z/z\right)^{\omega/2}
\eeq
brings it to the form
\beq
A_a=0\,,\quad \phi_0=\fr{i}{2y}\sum_iB_iH_i\,,\quad \varphi=\fr{1}{y}\sum_iB_i^{1/2}E^+_i\,.
\eeq
which is the required behaviour (\ref{principal}). So, to impose the boundary condition at $\sigma\to0$ we require that the functions $q_i$ approach at zero the model solution (\ref{model0}):
\beq
\sigma\rightarrow 0: \quad q_j=-2\log\sigma+\log B_j+\dots\,.
\eeq
Later it will be more convenient to work with a different set of variables defined as $\chi_i=\sum_j A^{-1}_{ij}q_j$. For these variables the boundary condition becomes
\beq
\sigma\rightarrow 0: \quad \ex^{-\chi_i}=\sigma^{B_i}\,\prod_k B_k^{-A^{-1}_{ik}}+\dots\,.\label{chilong}
\eeq
It is easy to see from the Toda equations that in fact {\it any} solution, for which $\ex^{-\chi_i}$ at zero vanish polynomially, will behave like (\ref{chilong}), so, finally, the boundary condition can be formulated simply as 
\beq
\sigma\to0:\quad \ex^{-\chi_i}\to 0\,.\label{bc0}
\eeq

The boundary condition along the line $r=0$ away from the origin, {\it i.e.} for $\sigma\to\infty$, comes from the requirement that the fields must be non-singular along this line. Then we have
\beq
\sigma\rightarrow\infty:\quad q_i=-2m_i\sigma+\log (4C_j)+{\rm O}(\ex^{-\sigma})\,,\quad m_i=\mathfrak{r}_i+1\,. \label{bc1}
\eeq
Here we introduced some constants $C_j$ which will be fixed from the boundary condition at zero. The exponential falloff of the residual terms in (\ref{bc1}) is expected from the general properties \cite{books} of the open Toda systems. In terms of variables $\chi_i$ the boundary condition is
\beq
\sigma\rightarrow\infty:\quad \chi_i=-2\lambda_i\sigma+\eta_i+{\rm O}(\ex^{-\sigma})\,,\label{bc11}
\eeq
where $\eta_i$ are appropriate combinations of the constants $C_j$, and $\lambda_i=\sum_j A^{-1}_{ij}m_j$.

\subsection{Solving the Toda system}
In this section we find the solution of the Toda equations with boundary conditions (\ref{bc0}), (\ref{bc1}). We do it in two steps. First we construct a solution starting from ``initial values'' at $\sigma\to\infty$ defined by the boundary condition (\ref{bc1}), for a given set of constants $C_j$. Then we tune these constants to match the second boundary condition (\ref{bc0}). For the first step of this procedure the result is in fact already known and described {\it e.g.} in \cite{Kostant}, but our derivation is very simple, so we decided to include it here.

We introduce useful notations for two Cartan elements, $\chi=\sum_i\chi_iH_i$ and $\hat{\omega}=\sum_i\lambda_iH_i$. In terms of notations of the previous subsection 
\beq
\hat{\omega}=\omega+\delta^\vee\,.
\eeq
Let $\Lambda_s$, $s=1,\dots {\rm rank}(\mathfrak{g})$, be the fundamental weights of the Lie algebra $\mathfrak{g}_\CC$, {\it i.e.} the highest weights of the fundamental representations $\rho_s$. We have
\beq
\fr{2\langle \Lambda_s,\alpha_i\rangle}{\langle\alpha_i,\alpha_i\rangle}=\delta_{is}\,,
\eeq
where angle brackets denote the Killing form on the coalgebra. We will denote the vectors in the particular representation by $|v\rangle$, in particular $|\Lambda_s\rangle$ is the heighest weight vector of unit norm in the representation $\rho_s$. Also we will not distinguish in notation the abstract elements of the Lie algebra and their evaluation in the representations. 

It has been shown in \cite{Mansfield} using very simple arguments that the value of a solution of the open Toda system at time $\sigma$ can be obtained from its values at different times $\tau$ using the formula
\beq
\ex^{-\chi_s(\sigma)}=\ex^{-\chi_s(\tau)}\langle\Lambda_s|\exp\left[(\tau-\sigma)\dot{\chi}(\tau)+\sqrt{-1}(\tau-\sigma)\sum_j\ex^{q_j(\tau)/2}(E^+_j+E^-_j)\right]|\Lambda_s\rangle\,.\label{main1}
\eeq
Our strategy will be to take the limit $\tau$ to infinity and to use the asymptotic formula (\ref{bc1}) to derive the functions $\chi_i(\sigma)$. In the limit $\tau\rightarrow\infty$ all the subleading terms in the exponent in (\ref{main1}) can be dropped, as will be clear from what follows. In this limit we obtain
\begin{flalign}
&\ex^{-\chi_s(\sigma)}=\lim_{\tau\to\infty}\ex^{2\lambda_s\tau-\eta_s}\langle\Lambda_s|\exp\left[2(-\tau+\sigma)\hat{\omega}+\tau\sum_j\ex^{-m_j\tau}\sqrt{-4C_j}(E^+_j+E^-_j)\right]|\Lambda_s\rangle\,\label{main2}\,.&
\end{flalign}
To calculate explicitly this limit we use the formula
\beq
\ex^{A+B}=\sum_m\int_0^1{\rm d}t_m\int_0^{t_m}{\rm d}t_{m-1}\dots\int_0^{t_2}{\rm d}t_1\ex^{(1-t_m)A}B\ex^{(t_m-t_{m-1})A}B\dots B\ex^{t_1A}\,.\label{polyakov}
\eeq
Operator $A$ will represent the Cartan part of the exponent in (\ref{main2}), and operator $B$ the $E^\pm$ part. The operators $B$ acting on $|\Lambda_s\rangle$ will prepare some weight state\footnote{First a particular term $E^+_{k_1}\dots E^-_{k_m}$ in the expansion of $B^n$ should be taken, for simplicity we don't write it explicitly here.}, and hence the operators $A$ in the exponents can be replaced by their appropriate eigenvalues $A_0,\dots A_m$. Then the integrals can be performed to give
\beq
\ex^{A+B}|\Lambda\rangle=\sum_{m=0}^\infty \sum_{k=0}^m \ex^{A_k}\fr{1}{\prod_{j\ne k}(A_k-A_j)}B\dots B|\Lambda\rangle \label{polyakov2}
\eeq
Consider a particular term in the expansion (\ref{polyakov2}) of (\ref{main2}) of the form
\beq
{\rm const}\times \ex^{2\lambda_s\tau}\langle\Lambda_s|\left(\prod_{i=1}^m E^+_{a_{i}}\right)\exp\left[(-\tau+\sigma)\sum_i2\lambda_iH_i\right]\exp\left[-2\tau\sum_{i=1}^mm_{a_i}\right]\left(\prod_{j=1}^m E^-_{b_j}\right)|\Lambda_s\rangle\,,\nonumber
\eeq
where $\{a_i\}$ is a set of $m$ indices, and $\{a_i\}$ is some permutation of it. It is easy to check that for a term of this form the exponentials of $\tau$ cancel (and so do the powers of $\tau$). Therefore the terms of any other form (where the Cartan exponential stands not in the middle of the string of $E$'s, or where $E^+$ and $E^-$ mix together on one side of the Cartan exponential) will be exponentially suppressed in $\tau$, because every extra insertion of $E^+_iE^-_i$ gives a factor $\exp(-2m_i\tau)$, and $m_i=\mathfrak{r}_i+1>0$.

Following this logic, the result can be written as follows. Let $\Delta_s$ be the set of weights of the fundamental representation $\rho_s$.  A weight $w\in \Delta_s$ of level $n(w)$ can be represented as
\beq
w=\Lambda_s-\sum_{l=1}^{n(w)}\alpha_{j_l}\,, \quad \alpha_i\in\Delta\,.
\eeq
Then we have from (\ref{main2}) and (\ref{polyakov2}):
\beq
\ex^{-\chi_s(\sigma)}=\ex^{-\eta_s}\sum_{w\in\Delta_s}\left[\exp\left(2\sigma w(\hat{\omega})\right)\langle v_w(\hat{\omega})|v_w(\hat{\omega})\rangle (-1)^{n(w)} \prod_{l=1}^{n(w)} C_{j_l}\right]\,.\label{long}
\eeq
Here the vector $|v_w(\hat{\omega})\rangle$ is defined as follows. Let ${\bf s}$ enumerate ways in which the weight space $w$ can be reached from the highest weight, i.e. each ${\bf s}$ corresponds to a sequence of the form 
\beq 
E_{j_{n(w)}}^-\dots E_{j_1}^-|\Lambda\rangle
\eeq
where again $\alpha_{j_{n(w)}}+\dots+\alpha_{j_1}=\Lambda-w$. On the way from $\Lambda$ to $w$ we encounter weights 
\beq
(\Lambda=w_1, w_2, \dots, w_{n(w)}, w_{n(w)+1}=w)\,.
\eeq
Then $|v_w(\hat{\omega})\rangle$ is a vector in the weight space of $w$: 
\beq
|v_w(\hat{\omega})\rangle=\sum_{\bf s} \prod_{a=1}^{n(w)} \frac{1}{w(\hat{\omega})-w_a(\hat{\omega})}E_{j_{n(w)}}^-\dots E_{j_1}^-|\Lambda\rangle\,,
\eeq

The next step is to fix the ${\rm rank}(\mathfrak{g})$ constants $C_i$ from the ${\rm rank}(\mathfrak{g})$ algebraic equations~(\ref{bc0}):
\beq
\sum_{w\in\Delta_s}\left[\langle v_w(\hat{\omega})|v_w(\hat{\omega})\rangle (-1)^{n(w)} \prod_{l=1}^{n(w)} C_{j_l}\right]=0\,.\label{algeqs}
\eeq
We conjecture that this solution is unique and is given by
\beq
C_i=\prod_{\beta_j\in\Delta_+}\left(\beta_j(\hat{\omega})\right)^{2\langle\alpha_i,\beta_j\rangle/\langle\beta_j,\beta_j\rangle}\,,\label{conj}
\eeq
where $\Delta_+$ is the set of positive roots. 

With this assumption, and using explicit expression for the constants $\eta_i$ in terms of $C_j$, the formula (\ref{long}) becomes
\begin{flalign}
&\ex^{-\chi_s(\sigma)}=2^{-B_s}\sum_{w\in\Delta_s}\left[\exp\left(2\sigma w(\hat{\omega})\right)\,\langle v_w(\hat{\omega})|v_w(\hat{\omega})\rangle\,(-1)^{n(w)}\,\prod_{\beta_a\in\Delta_+}\left(\beta_a(\hat{\omega})\right)^{-2\langle w,\beta_a\rangle/\langle\beta_a,\beta_a\rangle}\right]&\label{long1}
\end{flalign}

We have explicitly checked that (\ref{conj}) is indeed a solution of (\ref{algeqs}) for the algebras $B_2$ and $G_2$. The corresponding formulae are collected in the Appendix. In the next subsection we prove that it is also true for the $A_n$ algebras. We also believe that this solution is unique, but this has been checked only for $A_1$ and $A_2$.

\subsection{Proof of the conjecture for the case of $A_n$}
For the case of $A_n$ algebra the result (\ref{long1}) can be proved and further simplified. Idea of the proof comes from the following observation. Suppose we have solved the equations (\ref{algeqs}) and thus have the constants $C_j$ as functions of $\hat{\omega}$. The formula (\ref{long}) will then assume form
\beq
\ex^{-\chi_i(\sigma)}=\sum_{w\in\Delta_i}Q^i_w(\hat{\omega})\,\ex^{2\sigma w(\hat{\omega})}\,,\label{suppose}
\eeq
with some coefficients $Q_w(\hat{\omega})$ depending on $\hat{\omega}$. In our construction $\hat{\omega}$ was supposed to be positive, {\it i.e.} to lie in the positive Weyl chamber, but the r.h.s. of (\ref{suppose}) makes sense for any $\hat{\omega}$. It will still be solution of the Toda equations and will satisfy the boundary condition (\ref{bc0}). Suppose we substitute instead of positive $\hat{\omega}$ its Weyl transform $\omega_1=S\hat{\omega}$, $S\in W$. In the limit $\sigma\to\infty$ the behaviour of the function $\chi_i(\sigma)$ will be governed by the largest exponential, which is $(S^{\dagger-1}\Lambda_i)(\omega_1)=\Lambda_i(\hat{\omega})$, so these functions satisfy also the same boundary condition (\ref{bc11}) at infinity. Assuming uniqueness of the solution, we find that the expression (\ref{suppose}) must be Weyl-invariant:
\beq
Q^i_w(S\hat{\omega})=Q^i_{S^{\dagger}w}(\hat{\omega})\,.\label{inv}
\eeq
Thus, the strategy of the proof will be the following. First we check that the expression (\ref{long1}) is Weyl invariant. Then, using the fact that the fundamental representations of $A_n$ are minuscule, we obtain simple expressions for the constants $Q^i_w(\hat{\omega})$ by Weyl transformaions from the highest weight term. The resulting formula is simple enough to check the boundary condition (\ref{bc0}) directly.

Let $S_i\in W$ be the Weyl reflections along the simple roots. They generate the Weyl group, so we need to check invariance only under these reflections. Also, as we saw in the previous subsection, the solution is uniquely determined by the asymptotics at infinity, {\it i.e.} by the leading exponential in the sum (\ref{long}) and the coefficient in front of it. Among the simple Weyl reflections the only one which changes the leading exponential in the expression for $\chi_j(\sigma)$ is $S_j$. Moreover, this reflection exchanges the leading exponential with the next-to-leading one:
\beq
S_i: \Lambda_j\rightarrow \Lambda_j-\alpha_i \delta_{ij}\,.
\eeq
The coefficients in front of the two leading exponentials in (\ref{long1}) can be explicitly computed. They are
\beqn
&&Q^i_{\Lambda_i}(\hat{\omega})=2^{-B_i}\prod_{\beta_a\in\Delta_+}\left(\beta_a(\hat{\omega})\right)^{-2\langle \Lambda_i,\beta_a\rangle/\langle\beta_a,\beta_a\rangle}\,,\label{hwcoef}\\
&&Q^i_{\Lambda_i-\alpha_i}(\hat{\omega})=-\fr{2^{-B_i}}{(\alpha_i(\hat{\omega}))^2}\prod_{\beta_a\in\Delta_+}\left(\beta_a(\hat{\omega})\right)^{-2\langle \Lambda_i-\alpha_i,\beta_a\rangle/\langle\beta_a,\beta_a\rangle}\,.\nonumber
\eeqn
It is straightforward to check that these expressions satisfy the invariance condition (\ref{inv}) for the simple reflection $S_i$, and thus the whole expression (\ref{long1}) is Weyl-invariant.

Now we can use the fact that the fundamental representations of $A_n$ are minuscule to restore all the coefficients $Q^i_w(\hat{\omega})$, starting from the highest weight term. Concretely, if the weight $w\in\Delta_i$ can be obtained from the highest weight as $w=S\Lambda_i$, then
\beq
Q^i_w(\hat{\omega})=Q_{\Lambda_i}(S^\dagger\hat{\omega})\,.\label{mapping}
\eeq
To evaluate this explicitly we need a few facts. First, if a weight $w$ lies in a string along the positive root $\beta$
\beq
w-r\beta,\dots, w,\dots, w+p\beta
\eeq
then $p-r=2\langle w,\beta\rangle/\langle\beta,\beta\rangle$. If $w$ is an image of the heighest weight $\Lambda$ under the Weyl reflection $S$, then for each $\beta$ either $r=0$, or $p=0$, because Weyl group maps strings into strings. If $\beta$ is mapped to a positive root, then $p=0$, if it is mapped to a negative root, then $r=0$. It is also easy to see that the total length of strings that start at the highest weight $\Lambda_i$ is equal to $B_i$.

The meaning of the formula (\ref{hwcoef}) now becomes clear. If the highest weight lies in strings along $\beta_a$ of length $r_a$, then
\beq
Q^i_{\Lambda_i}(\hat{\omega})=\prod_{\beta_a\in\Delta_+}\left(\beta_a[2\hat{\omega}]\right)^{-r_a}\,.
\eeq
Now it should be also clear how the mapping (\ref{mapping}) works. For a given weight $w$ let $\Delta^+(w)$ be the set of positive roots which can be subtracted from $w$, {\it i.e.} for which $r\ne0$, and $\Delta^-(w)$ be the set of positive roots which can be added to $w$, {\it i.e.} with $p\ne 0$. Then
\beq
Q^i_w(\hat{\omega})=\prod_{\beta_a\in\Delta^+(w)}(\beta_a[2\hat{\omega}])^{-r_a}\prod_{\beta_b\in\Delta^-(w)}(-\beta_b[2\hat{\omega}])^{-p_a}\,.
\eeq
We note that for the fundamental representations of $A_n$ all the strings are two elements long. It can be seen from the explicit matrix expressions for the roots and weights. The fundamental weight $\Lambda_j$ is
\beq
\Lambda_j={\rm diag}\Bigl(\,\underbrace{-\fr{j}{n+1}+1,\,\dots,\,-\fr{j}{n+1}+1,}_{j}\,\underbrace{-\fr{j}{n+1},\,\dots,\,-\fr{j}{n+1}}_{n+1-j}\,\Bigr)\,,
\eeq
the other weights in the same fundamental representation differ by permutations of the diagonal elements, the positive roots as usual have $1$ and $-1$ on the diagonal, and the Killing form is simply the matrix trace. We see that for any positive root the number $2\langle\Lambda_j,\beta\rangle/\langle\beta,\beta\rangle$ is either $1$ or $0$, hence all the strings are two elements long. Finally we can write a simplified version of the formula for the functions $\chi_i(\sigma)$ in the case of $A_n$ Lie algebra:
\beq
\ex^{-\chi_j(\sigma)}=\sum_{w\in\Delta_j}\ex^{\sigma\,w[2\hat{\omega}]}\biggl(\prod_{\beta_a\in\Delta^+(w)}\beta_a[2\hat{\omega}]\prod_{\beta_b\in\Delta^-(w)}(-\beta_b[2\hat{\omega}])\biggr)^{-1}\,.\label{simple}
\eeq

Having this formula, we can prove that these functions indeed satisfy the boundary condition (\ref{bc0}). We need to prove that the r.h.s. of this expression at $\sigma=0$ is zero. It is a rational function of $\hat{\omega}$ which goes to zero at $\hat{\omega}\to\infty$, and potentially has poles when for some positive root, {\it e.g.} $\beta_1$, we have $\beta_1(\hat{\omega})\to 0$. As all the strings are two terms long, the poles come in pairs. Suppose that such pair of poles comes from the terms corresponding to weights $w$ and $w-\beta_1$, {\it i.e.} assume $\beta_1\in \Delta^+(w)$ and $\beta_1\in \Delta^-(w-\beta_1)$. Looking at the matrix representation of weights and computing the numbers $r_a, p_a$ for $w$ and $w-\beta_1$ it is not difficult to verify that if $\beta_a$, $a\ne1$, appears in $\Delta^\pm(w)$, then $\beta_a$ or $\beta_a\pm\beta_1$ will appear in $\Delta^\pm(w-\beta_1)$. Then in the limit $\beta_1(\hat{\omega})\to0$ we get $Q^i_w(\hat{\omega})+Q^i_{w-\beta_1}(\hat{\omega})\to 0$. Thus, the expression (\ref{simple}) at $\sigma=0$ is a rational function of $\hat{\omega}$ with no poles, and hence is zero.

\section{Solution with a line singularity}
In this section we construct solutions of the Bogomolny equations for a boundary surface operator living in the $t,y$ plane at $z=0$. In this case along the line $z=0$ the fields have a singularity of the form
\beqn
A=\alpha{\rm d}\theta\,,\quad \varphi=\mu \fr{{\rm d}z}{z} \label{neabsing}
\eeqn
with parameters $\alpha\in i\mathfrak{h}$ and $\mu\in\mathfrak{h}_\CC$. Here $\theta$ is the polar angle in the $z$-plane. We take the solution to equations (\ref{commut}) to be
\beq
\varphi_0(z)=\fr{\mu}{z}\,.
\eeq
Substituting this ansatz for $\varphi_0$ to equation (\ref{fullmt}) and assuming again conformal invariance, we get
\beq
\pt_\sigma\left(\pt_\sigma h h^{-1}\right)+\left[\mu^\dagger, h\mu h^{-1}\right]=0\,.\label{natoda}
\eeq
Unlike the case of line operator, here an abelian ansatz for the gauge transformation does not work. What we get is a version of non-abelian Toda equation.

The boundary conditions at $\sigma\rightarrow \infty$ are defined by (\ref{neabsing}). In terms of gauge transformation $g$ they are
\beqn
&&g\mu g^{-1}=\mu+\dots\,,\nnr
&&g^{\dagger-1}\pt_\sigma hh^{-1}g^\dagger=2i\alpha+\dots\,.\label{bcinfty}
\eeqn
On the plane $y=0$, or at $\sigma\rightarrow 0$, the boundary conditions are the same as for the line operator. For the gauge transformation they read as
\beqn
&&g\mu g^{-1}=\fr{1}{\sigma}\sum_i\sqrt{B_i} E^+_i+\dots\,,\nnr
&&g^{\dagger-1}\pt_\sigma hh^{-1}g^\dagger=-\fr{1}{\sigma}\sum_i B_iH_i+\dots\,.
\eeqn

The non-abelian Toda equation (\ref{natoda}) is a degenerate case of a more general non-abelian Toda system studied in the literature \cite{DubrKrichever}.  General non-abelian Toda equations were first considered by A.~Polyakov. They have form
\beq
\pt_t(\pt_t h_n h_n^{-1})=h_{n-1}h_n^{-1}-h_n h_{n+1}^{-1}\,,
\eeq
where $h_n$ are some matrices. These equations have a Lax representation, and thus the solutions can be constructed in a standard way in terms of $\theta$-functions on the spectral curve, see Appendix by I.~Krichever in \cite{DubrKrichever}. Our equation corresponds to the case when there is only one node with twisted periodicity condition $h_{n-1}=\mu^\dagger h_n\mu$.  For a general review of algebro-geometric methods of solving non-linear equations see \cite{DubrKrichever,DKN,Krichever}. 
We will sketch the technics as adopted to our reduced equation.

The equation (\ref{natoda}) can be written in the form of a zero-curvature condition
\beq
[\pt_\sigma-A(w),L(w)]=0\,.\label{flat}
\eeq
with a Lax pair
\beqn
&&L(w)=-\pt_\sigma hh^{-1}-wh\mu h^{-1}+\fr{1}{w}\mu^\dagger\,,\\
&&A(w)=\fr12 \pt_\sigma hh^{-1}-\fr12w h\mu h^{-1}-\fr12\fr{1}{w}\mu^\dagger\,,
\eeqn
where $w\in\CC$ is a spectral parameter. 

Let $\Psi(\sigma,w)$ be an eigenvector of the Lax operator in some representation, with eigenvalue $E$:
\beq
L(w)\Psi(\sigma,w)=E\Psi(\sigma,E)\,.\label{psieq1}
\eeq
and let it also satisfy equation
\beq
(\pt_\sigma-A(w))\Psi(\sigma,E)=0\,.\label{psieq2}
\eeq
This object is called the Baker-Akhiezer function (in our case it is a vector function). The commutativity condition (\ref{flat}) ensures that such function exists. From the usual argument it follows that $\Psi$ is a well-defined function on the spectral curve
\beq
\Sigma:\quad \det(L(w)-E)=0\,.\label{spcurve}
\eeq
To find the solutions of our equation we need to construct the spectral curve and the Baker-Akhiezer function.

The spectral curve is invariant under the Hamiltonian flow. It depends only on the conserved quantities characterizing the solution. Therefore we can use the asymptotics at $\sigma\to\infty$ to find its explicit form. From (\ref{bcinfty}) we get
\beq
L(w,\sigma\to\infty)=-2i\alpha-w\mu+\fr{1}{w}\mu^\dagger\,.\label{Lasympt}
\eeq
The method can be used for any matrix Lie algebra, but for simplicity we now specialize to the case of $A_N$. So, the Cartan elements become diagonal matrices,
\beqn
-i\alpha&=&\diag(\nu_1\,\dots,\nu_N)\,,\nnr
\mu&=&\diag(\mu_1,\dots,\mu_N)\,.
\eeqn
For simplicity we will assume that $\mu_i\ne0$ and also that $\mu$ is a regular element, so $\mu_i\ne\mu_j$ for $i\ne j$.

From the asymptotic expression (\ref{Lasympt}) for the Lax operator the spectral curve $\Sigma$ can be found to be a union of $N$ sheets
\beqn
{\rm j}^{\rm th}\,{\rm sheet\,}\Sigma_j:\,\,&E=-\mu_j w+\bar{\mu}_j\fr{1}{w}+2\nu_j\,.&
\eeqn
Each pair of sheets $(\Sigma_i,\Sigma_j)$ is joined at two points $w_{ij}$, $w_{ji}$:
\beq
w_{ij}=\fr{\nu_i-\nu_j+\omega_{ij}}{\mu_i-\mu_j}\,,
\eeq
where we denoted
\beq
\omega_{ij}=\sqrt{(\nu_i-\nu_j)^2+|\mu_i-\mu_j|^2}\,.
\eeq
On each sheet of the curve there are two marked points
\beqn
&{\rm P}^+_j:\quad w\rightarrow \infty\,,\,\, &E=-\mu_jw+2\nu_j+\dots\,,\nnr
&{\rm P}^-_j:\quad w\rightarrow 0\,,\,\, &E=\bar{\mu}_j\fr{1}{w}+2\nu_j+\dots\,.
\eeqn

Now we want to reconstruct the Baker-Akhiezer function on the curve $\Sigma$. To this end we will prescribe its analytical properties and then show that it is indeed a solution of (\ref{psieq1}), (\ref{psieq2}). The function $\Psi$ is a meromorphic function away from the marked points $P^\pm_j$, with $N(N-1)$ poles which are independent of $\sigma$. At the marked points we require the following asymptotics:
\beq
\Psi|_{P^\pm_j}=\ex^{\pm E\sigma/2}\left(\xi^\pm_j+\fr{1}{E}\eta^\pm_j+O(1/E^2)\right)\,,\label{psiasympt}
\eeq
with normalization $\xi_j^-=v_j$, where $v_j$ are the basis vectors, $(v_j)_k=\delta_{jk}$.

Since the curve is degenerate, the Baker-Akhiezer function is in fact a collection of functions on each sheet 
\beq
\Psi_j=\Psi|_{\Sigma_j}
\eeq
satisfying the gluing equations on the intersections
\beq
\Psi_i(w_{ij})=\Psi_j(w_{ij})\,.\label{gluing}
\eeq
Let $\Psi_j$ have poles $w_j^1,\dots,\,w_j^{n_j}$. The asymptotic conditions (\ref{psiasympt}) and analyticity fix the form of the function,
\beq
\Psi_j=\exp\left(-(\mu_jw+\fr{\bar{\mu}_j}{w})\frac{\sigma}{2}\right)\frac{\ex^{\nu_j\sigma}\xi^+_j w^{n_j}+a_j^{n_j-1}w^{n_j-1}+\dots+a_j^{1}w+\ex^{-\nu_j\sigma}\prod_{\alpha=1}^{n_j}(-w_j^\alpha)v_j}{\prod_{\alpha=1}^{n_j}(w-w_j^\alpha)}\,,
\eeq
where $a_j^{1},\dots,\,a_j^{n_j-1}$ are unknown vector coefficients depending on $\sigma$. So, for fixed poles $w_j^\alpha$ each $\Psi_j$ depends on $N\times n_j$ unknown parameters, which are vectors $a_j^\alpha$ and $\xi^+_j$. The total number of parameters for $\Psi$ is $N^2\sum n_j=N^2(N-1)$, and it is equal to the number of equations (\ref{gluing}). It follows that the function $\Psi$ is uniquely defined by its analytical properties.

Let us prove that $\Psi$ is indeed a solution of the Lax equations (\ref{psieq1}), (\ref{psieq2}). Consider the function $(L(w)-E)\Psi$. Away from the marked points it has the same properties as $\Psi$. At $P^-_j$, where $w\rightarrow0$, one can compute the asymptotics using (\ref{psiasympt}):
\beq
(L(w)-E)\Psi|_{P^-_j}=\ex^{-E\sigma/2}\left(\fr{1}{w}(\mu^\dagger-\bar{\mu}_j)v_j+\fr{\mu^\dagger-\bar{\mu}_j}{\bar{\mu}_j}\eta^-_j-(\pt_\sigma hh^{-1}+2\nu_j)v_j+O(w)\right)\,.\nonumber
\eeq
One can immediately see that the term of order $1/w$ is zero. We can also cancel the order one term by taking the momentum term $\pt_\sigma hh^{-1}$ in the Lax operator from the equation\footnote{Note that for the diagonal components it gives $(\pt_\sigma hh^{-1})_{jj}=-2\nu_j$, which is the asymptotic value at $\sigma\rightarrow\infty$. This is consistent with the Toda equation, because one can see that the Cartan part of $\pt_\sigma hh^{-1}$ is constant by that equation.}
\beq
\fr{\mu^\dagger-\bar{\mu}_j}{\bar{\mu}_j}\eta^-_j=(\pt_\sigma hh^{-1}+2\nu_j)v_j\,.
\eeq
The asymptotics at $P^+_j$, where $w\rightarrow\infty$, are
\beqn
&&(L(w)-E)\Psi|_{P^+_j}=\nnr
&&\ex^{E\sigma/2}\left(w(\mu_j-h\mu h^{-1})\xi_j^++\fr{h\mu h^{-1}-\mu_j}{\mu_j}\eta_j^+-(\pt_\sigma hh^{-1}+2\nu_j)\xi_j^++O(\fr{1}{w})\right)\,.\nonumber
\eeqn
The $w$ order term is canceled by choosing $h$ in the Lax operator so that $\xi^+_j=hv_j$. We see that the function $(L(w)-E)\Psi$ has the same analytical properties as $\Psi$, except that it behaves better at points $P^j_-$. By uniqueness of $\Psi$ we conclude that $(L(w)-E)\Psi=0$. Similar arguments show that $(\pt_\sigma-A)\Psi$ also vanishes, so we have proved that the function $\Psi$ reconstructed from the analytical properties indeed satisfies the Lax equations (\ref{psieq1}), (\ref{psieq2}). On the way we also found that the solution $h(\sigma)$ to the Toda equation can be inferred from the asymptotics at $P^+_j$, where $\xi^+_j=hv_j$.

To find the vectors $\xi^+_j$, and hence the solution, we need to solve the equations (\ref{gluing}), which are $N$ systems of $N(N-1)$ linear equations each:
\beqn
&&\fr{\ex^{-\omega_{ij} \sigma/2}}{\prod_{\alpha=1}^{n_i}(w-w_i^\alpha)}\left(\hat{h}_{ki}w_{ij}^{N-1}+(a_{i,N-2})_k+\dots+(a_{i,1})_k w_{ij}+\delta_{ki}\prod_{\alpha=1}^{n_i}(-w_i^\alpha)\right)=\nnr
&&\fr{\ex^{\omega_{ij} \sigma/2}}{\prod_{\alpha=1}^{n_j}(w-w_j^\alpha)}\left(\hat{h}_{kj}w_{ij}^{N-1}+(a_{j,N-2})_k+\dots+(a_{j,1})_k w_{ij}+\delta_{kj}\prod_{\alpha=1}^{n_j}(-w_j^\alpha)\right)\,.\label{fullsystem}
\eeqn
Here we introduced a new variable $\hat{h}=\ex^{\nu\sigma}h\ex^{\nu\sigma}$. The poles $w_j^\alpha$ should be then fixed from the boundary condition at $\sigma\rightarrow0$, which we have not yet imposed. We will not do this directly, but rather will use some intuition gained from the abelian case discussed in the previous sections. There we have seen that the boundary conditions at $\sigma\rightarrow0$ mean essentially that the fields have singularity, and $\it any$ solution, for which all components are singular at $\sigma\rightarrow 0$, will automatically satisfy the boundary condition.  In the non-abelian case we expect something similar to happen. The $N(N-1)$ by $N(N-1)$ matrices of the equations (\ref{fullsystem}) must become degenerate at the origin for the solution to have a singularity. We make a guess that the singularity that we need will occur if these matrices have the minimal possible rank at the origin, which is one. This will happen if we put equal number of poles on each sheet, and the poles at different sheets will lie above each other, {\it i.e.} $w_i^\alpha=w_j^\alpha$ for any $i$ and $j$. Then the equations (\ref{fullsystem}) take form
\beqn
&&\ex^{-\omega_{ij} \sigma/2}\left(\hat{h}_{ki}w_{ij}^{N-1}+(a_{i,N-2})_k+\dots+(a_{i,1})_k w_{ij}+\delta_{ki}\right)=\nnr
&&\ex^{\omega_{ij} \sigma/2}\left(\hat{h}_{kj}w_{ij}^{N-1}+(a_{j,N-2})_k+\dots+(a_{j,1})_k w_{ij}+\delta_{kj}\right)\,.
\eeqn
The poles have canceled out from the equations. The multipliers $\prod_{\alpha=1}^{N-1}(-w^\alpha)$ were absorbed by rescaling of $\hat{h}$. 

The equations can be reformulated in a slightly more compact way. Define a $w$-dependent operator which is an $N$ by $N$ matrix
\beq
\mathcal{O}(w)=\hat{h}w^{N-1}+a_{N-2}w^{N-2}+\dots+a_1w+{c}\,.\label{operator}
\eeq
Then the linear equations can be formulated as follows. Given a set of vectors 
\beq 
u_{ij}=\ex^{-\omega_{ij}\sigma/2}v_i-\ex^{\omega_{ij}\sigma/2}v_j
\eeq
sitting at points $w_{ij}$, find the operator which defines this ``skyscraper sheaf'' by
\beq
\mathcal{O}(w)u=0\,.\label{sheaf}
\eeq
This operator $\mathcal{O}(w)$ is, up to a simple twist, the P-exponent of the Lax connection $A(w)$. 

The letter ${c}$ appearing in the definition (\ref{operator}) is related to a following subtlety. Our procedure gives matrix $\hat{h}$ only up to multiplication by diagonal matrices on the left and on the right, and the matrix that we get is in general not hermitian. This can be remedied by introducing some diagonal matrix ${c}$ in  (\ref{operator}), which must be fixed from the requirement of the hermiticity of $\hat{h}$.

Let us consider an example. For $N=2$ the equations (\ref{sheaf}) become
\beq
\left(\hat{h}w_{ij}+c\right)u_{ij}=0\,,
\eeq
saying that $u_{12}$ and $u_{21}$ are the eigenvectors of $c^{-1}\hat{h}$. Then one immediately obtains
\beqn
h=\fr{1}{|\mu|}\left(
\begin{array}{cc}
\ex^{-2\nu\sigma}\left(\nu+\fr{\omega}{2}\coth\omega\sigma\right) & \fr{\omega}{2}\fr{1}{\sinh\omega\sigma}  \vspace{5pt} \\
\fr{\omega}{2}\fr{1}{\sinh\omega\sigma} & \ex^{2\nu\sigma}\left(-\nu+\fr{\omega}{2}\coth{\omega\sigma}\right)
\end{array}\right)\,.
\eeqn
Unfortunately, already for $N=3$ the solution becomes very clumsy, and we were unable to simplify it. But we have checked numerically that our guess about the boundary conditions at $\sigma\rightarrow0$ is correct for $N=3$ as well.

\section*{Acknowledgments}
I would like to thank E.~Witten for many helpful discussions and for sharing unpublished notes on the generalized Bogomolny equations. I also thank A.~Gorsky, B.~Basso and A.~Zhiboedov for useful discussions.

\section*{Appendix}
In this Appendix we collect some explicit formulae for the 't~Hooft operator solution. Let us recall the notations. The operator is labeled by cocharacter $\omega\in\Gamma^\vee_{ch}$. Another useful variable is $\hat{\omega}=\omega+\delta^\vee$. Let $\Delta$ be the set of simple roots $\alpha_i$, then $\alpha_i(\hat{\omega})=m_i$. Let $E_\alpha$ be the raising generators corresponding to the simple roots. Then the fields on the solution are
\beqn
&&\varphi=\fr{1}{r}\sum_{\alpha\in\Delta}\exp\left[\alpha(i\omega\theta+\fr12\chi(\sigma))\right]E_\alpha\,,\nnr
&&\phi_0=-\fr{i}{2\rho}\pt_\sigma\chi(\sigma)\,,\nnr
&&A=-i\left(\hat{\omega}+\fr12\frac{y}{\sqrt{y^2+r^2}}\pt_\sigma\chi(\sigma)\right){\rm d}\theta\,.\nonumber
\eeqn
Here $\chi(\sigma)=\sum \chi_i(\sigma)H_i$\,, and the functions $\chi_i(\sigma)$ are collected below for the algebras $A_1$, $A_2$, $A_3$, $B_2$ and $G_2$.
\vspace{5mm}
\\${\bf A_1}$
\begin{flalign*}
\exp(-\chi)=\fr{\sinh(m\sigma)}{m}\,.&&
\end{flalign*}
${\bf A_2}$
\begin{flalign*}
&\exp(-\chi_1)=\fr14\left(\fr{\exp\left[\fr23\sigma(2m_1+m_2)\right]}{m_1(m_1+m_2)}-\fr{\exp\left[\fr23\sigma(-m_1+m_2)\right]}{m_1 m_2}+\fr{\exp\left[-\fr23\sigma(m_1+2m_2)\right]}{m_2(m_1+m_2)}\right)\,,&\\
&\exp(-\chi_2)=\fr14\left(\fr{\exp\left[\fr23\sigma(m_1+2m_2)\right]}{m_2(m_1+m_2)}-\fr{\exp\left[\fr23\sigma(m_1-m_2)\right]}{m_1m_2}+\fr{\exp\left[-\fr23\sigma(2m_1+m_2)\right]}{m_1(m_1+m_2)}\right)\,.&
\end{flalign*}
${\bf A_3}$
\begin{flalign*}
\exp(-\chi_1)=&\fr18\left(\fr{\exp\left[\fr12\sigma(3m_1+2m_2+m_3)\right]}{m_1(m_1+m_2)(m_1+m_2+m_3)}-\fr{\exp\left[\fr12\sigma(-m_1+2m_2+m_3)\right]}{m_1m_2(m_2+m_3)}+\right.&\\
&\left.\fr{\exp\left[\fr12\sigma(-m_1-2m_2+m_3)\right]}{m_2m_3(m_1+m_2)}-\fr{\exp\left[-\fr12\sigma(m_1+2m_2+3m_3)\right]}{m_3(m_2+m_3)(m_1+m_2+m_3)}\right)\,,&\\
\exp(-\chi_2)=&\fr{1}{8}\left(\fr{\cosh\left[\sigma(m_1+2m_2+m_3)\right]}{m_2(m_1+m_2)(m_2+m_3)(m_1+m_2+m_3)}-\fr{\cosh\left[\sigma(m_1+m_3)\right]}{m_1m_2m_3(m_1+m_2+m_3)}+\right.&\\
&\left.\fr{\cosh\left[\sigma(m_1-m_3)\right]}{m_1m_3(m_1+m_2)(m_2+m_3)}\right)\,,&\\
\exp(-\chi_3)=&\fr18\left(\fr{\exp\left[\fr12\sigma(m_1+2m_2+3m_3)\right]}{m_3(m_2+m_3)(m_1+m_2+m_3)}-\fr{\exp\left[\fr12\sigma(m_1+2m_2-m_3)\right]}{m_2m_3(m_1+m_2)}+\right.&\\
&\left.\fr{\exp\left[\fr12\sigma(m_1-2m_2-m_3)\right]}{m_1m_2(m_2+m_3)}-\fr{\exp\left[-\fr12\sigma(3m_1+2m_2+m_3)\right]}{m_1(m_1+m_2)(m_1+m_2+m_3)}\right)\,.&
\end{flalign*}
${\bf B_2}$
\begin{flalign*}
\exp(-\chi_1)=&\fr{1}{8}\left(\fr{\cosh\left[2\sigma(m_1+m_2)\right]}{m_1(m_1+m_2)^2(m_1+2m_2)}-\fr{\cosh\left[2\sigma m_2\right]}{m_1 m_2^2(m_1+2m_2)}+\fr{1}{m_2^2(m_1+m_2)^2}\right)\,,&\\
\exp(-\chi_2)=&\fr{1}{4}\left(\fr{\sinh\left[\sigma(m_1+2m_2)\right]}{m_2(m_1+m_2)(m_1+2m_2)}-\fr{\sinh\left[\sigma m_1\right]}{m_1 m_2 (m_1+m_2)}\right)\,.&
\end{flalign*}
${\bf G_2}$
\begin{flalign*}
\exp(-\chi_1)=&\fr{1}{512}\left(\fr{\cosh\left[2\sigma(2m_1+3m_2)\right]}{m_1(m_1+m_2)^3(m_1+2m_2)^3(m_1+3m_2)(2m_1+3m_2)^2}-\right.&\\
&\fr{\cosh\left[2\sigma(m_1+3m_2)\right]}{m_1m_2^3(m_1+2m_2)^3(m_1+3m_2)^2(2m_1+3m_2)}+&\\
&\fr{3\cosh\left[2\sigma(m_1+2m_2)\right]}{m_2^3(m_1+m_2)^3(m_1+2m_2)^2(m_1+3m_2)(2m_1+3m_2)}-&\\
&\fr{3\cosh\left[2\sigma(m_1+m_2)\right]}{m_1m_2^3(m_1+m_2)^2(m_1+2m_2)^3(2m_1+3m_2)}+&\\
&\fr{3\cosh\left[2\sigma m_2\right]}{m_1m_2^2(m_1+m_2)^3(m_1+2m_2)^3(m_1+3m_2)}+&\\
&\fr{\cosh\left[2\sigma m_1\right]}{m_1^2m_2^3(m_1+m_2)^3(m_1+3m_2)(2m_1+3m_2)}-&\\
&\left.\fr{12(m_1^2+3m_1m_2+3m_2^2)}{m_1^2m_2^2(m_1+m_2)^2(m_1+2m_2)^2(m_1+3m_2)^2(2m_1+3m_2)^2}\right)\,,&\\
\exp(-\chi_2)=&\fr{1}{32}\left(\fr{\cosh\left[2\sigma(m_1+2m_2)\right]}{m_2(m_1+m_2)(m_1+2m_2)^2(m_1+3m_2)(2m_1+3m_2)}-\right.&\\
&\fr{\cosh\left[2\sigma(m_1+m_2)\right]}{m_1m_2(m_1+m_2)^2(m_1+2m_2)(2m_1+3m_2)}+&\\
&\left.\fr{\cosh\left[2\sigma m_2\right]}{m_1m_2^2(m_1+m_2)(m_1+2m_2)(m_1+3m_2)}-\fr{1}{m_2^2(m_1+m_2)^2(m_1+2m_2)^2}\right)\,.&
\end{flalign*}

\enddocument